\documentclass[aps,preprint]{revtex4-1}
\usepackage{graphicx}
\usepackage{amssymb}
\usepackage{amsmath}
\usepackage{color}
\usepackage{enumerate}
\usepackage{bm}
\begin{document}

\title{Angle Selective Piezoelectric Strain Controlled Magnetization Switching in Artificial Spin Ice Based Multiferroic System}

\author{Avinash Chaurasiya}
\altaffiliation{These authors contributed equally to this work}
\affiliation{Natural Sciences and Science Education, National Institute of Education, Nanyang Technological University, 637616, Singapore}

\author{Manish Anand}
\altaffiliation{These authors contributed equally to this work}
\affiliation{Department of Physics, Bihar National College, Patna University, Patna-800004, India.}

\author{Rajdeep Singh Rawat}
\email{rajdeep.rawat@nie.edu.sg}
\affiliation{Natural Sciences and Science Education, National Institute of Education, Nanyang Technological University, 637616, Singapore}

\date{\today}

\begin{abstract}

The prospect of all electrically controlled writing of ferromagnetic bits is highly desirable for developing scalable and energy-efficient spintronics devices. In the present work, we perform micromagnetic simulations to investigate the electric field-induced strain mediated magnetization switching in artificial spin ice (ASI) based multiferroic system, which is proposed to have a significant decrease in Joule heating losses compared to electric current based methods. As the piezoelectric strain-based system cannot switch the magnetization by $180^\circ$ in ferromagnets, we propose an ASI multiferroic system consisting of the peanut-shaped nanomagnets on ferroelectric substrate with the angle between the easy axis and hard axis of magnetization less than $90^\circ$. Here the piezoelectric strain-controlled magnetization switching has been studied by applying the electric field pulse at different angles with respect to the axes of the system. Remarkably, magnetization switches by $180^\circ$ only if the external electric field pulse is applied at some specific angles, close to the anisotropy axis of the system ($\sim 30^\circ-60^\circ$). Our detailed analysis of the demagnetization energy variation reveals that the energy barrier becomes antisymmetric in such cases, facilitating
the complete magnetization reversal. Moreover, we have also proposed a possible magnetization reversal mechanism with two sequential electric field pulses of relatively smaller magnitude. We believe that the present work could pave the way for future ASI-based multiferroic system for scalable magnetic field-free low power spintronics devices.



\end{abstract}
\maketitle

\section{Introduction}
There has been significant interest in electrically controlled information writing using artificial multiferroic to develop scalable energy-efficient spintronics devices~\cite{bibes2012,khanas2020,ustinov2019,chavez2018,trassin2015}. Various methods such as current-induced spin-transfer torque (STT), external magnetic field (using electromagnet), etc., are employed to achieve the magnetization switching in such a system~\cite{kwak2018,mu2006,manchon2019,sheng2018}. However, these are still facing challenges due to requirement of resource intensive magnetic field generation system and the energy loss in the form of Joule heating in current driven STT system. Therefore, developing magnetic memory with electric write and magnetic read is one of the highly desirable approaches for ultralow-power spintronics~\cite{chiba2008,fiebig2016,zhang2012}. Magnetoelectric-multiferroic systems consisting of more than one ferroic order are important material choices with the ability to control the magnetization using dual stimulus~\cite{biswas2017,cai2017,cherifi2014,chu2008,guo2018,chaurasiya2020,heron2011,yan2019}. However, using single-phase multiferroic materials is challenging due to weak magneto-electric coupling at room temperature~\cite{zhao2006}. Therefore, magneto-electric multiferroic heterostructure consisting of independent ferromagnet and ferroelectric have been explored as a promising candidate because of the significant magneto-electric coupling well above at the room temperature~\cite{li2017,liang2019}.
Magnetization manipulation has been realized through different routes, including charge coupling, exchange coupling, strain coupling, etc.~\cite{cuellar2014,wu2010,lei2013,zheng2007}. Among these, magnetization switching through strain-mediated magneto-electric coupling is demanding due to the long-range transfer of piezoelectric strain. Theoretically, it is proposed
that utilizing piezoelectric strain allows one to write information at ultralow energy of the order of aJ/bit~\cite{d2016}.
In recent years, the nanoscale patterned ferromagnet coupled with the ferroelectric has been the subject of great interest to control the magnetization through piezoelectric strain~\cite{cui2017,roy2013,wang2014,cui2013}. Fundamentally nanomagnets offer scalable, stable and simple magnetic states, making them an ideal candidate for the studies of magnetization dynamics~\cite{lendinez2019,porro2019}. Several attempts for magnetization switching in the patterned nanomagnetic system have been made utilizing artificial multiferroic heterostructure~\cite{buzzi2013,salehi2017}. 
However, the challenge of achieving piezoelectric strain-controlled magnetization switching in multiferroic artificial spin ice systems (ASI) is still unexplored.

We performed micromagnetic simulations to study the electric field controlled magnetization switching in a square ASI system comprising of peanut-shaped nanomagnets coupled to ferroelectric substrate in the present work. Conventionally, it is impossible to achieve the complete magnetization reversal using piezoelectric strain alone in such a case due to quadratic dependence of magnetization $M$ on strain $\epsilon$.
We design a square ASI system using the peanut-shaped nanomagnet instead of a conventionally used elliptical-shaped nanomagnet to overcome this issue, as shown in Fig.~(\ref{figure1}). The use of an engineered peanut-shaped nano-FMs provides an elegant en route to modify the angle between the easy and hard axis of magnetization (less than $90^\circ$)~\cite{cui2017}. In the elliptical-shaped nanomagnet, the easy and hard axes are orthogonal to each other, i.e. $90^\circ$~\cite{keswani2019}. It leads to impossibility of achieving the magnetization switching through piezoelectric strain in elliptical nanomagnets due to the uniaxial nature of strain which can provide the rotation of magnetization maximum by $90^\circ$. We designed an ASI system using a peanut-shaped nanomagnet with a unique easy axis and the smaller value of switching field to achieve deterministic magnetization switching. Here complete magnetization switching has been realized by applying the piezoelectric strain or an electric field at some specific angles with respect to the system's axes. The origin of angle specific magnetization switching is investigated by analyzing the variation of the demagnetization energy with the angle of the applied magnetic field for various piezoelectric strains rigorously. Our detailed analysis reveals that the energy barrier becomes antisymmetric in such
cases because of the demagnetization field, facilitating the complete magnetization reversal. Further, we have also proposed a possible magnetization reversal mechanism with two sequential electric field pulses of relatively smaller magnitude. It could be an efficient and alternative way to achieve the magnetization switching in ferromagnets with a weak magnetostriction effect. 
Therefore, observations made in the present work are highly desirable for future scalable magnetic field-free low power spintronics devices.

The rest of the article is organized as follows. In Sec. II, we  present the proposed ASI system and discuss the various energy terms and methods of simulations. We discuss simulations results in Sec. III. Finally, the summary and conclusion of the main results are provided in Sec. IV.
\section{Model}
\begin{figure}[!htb]
\centering\includegraphics[scale=0.10]{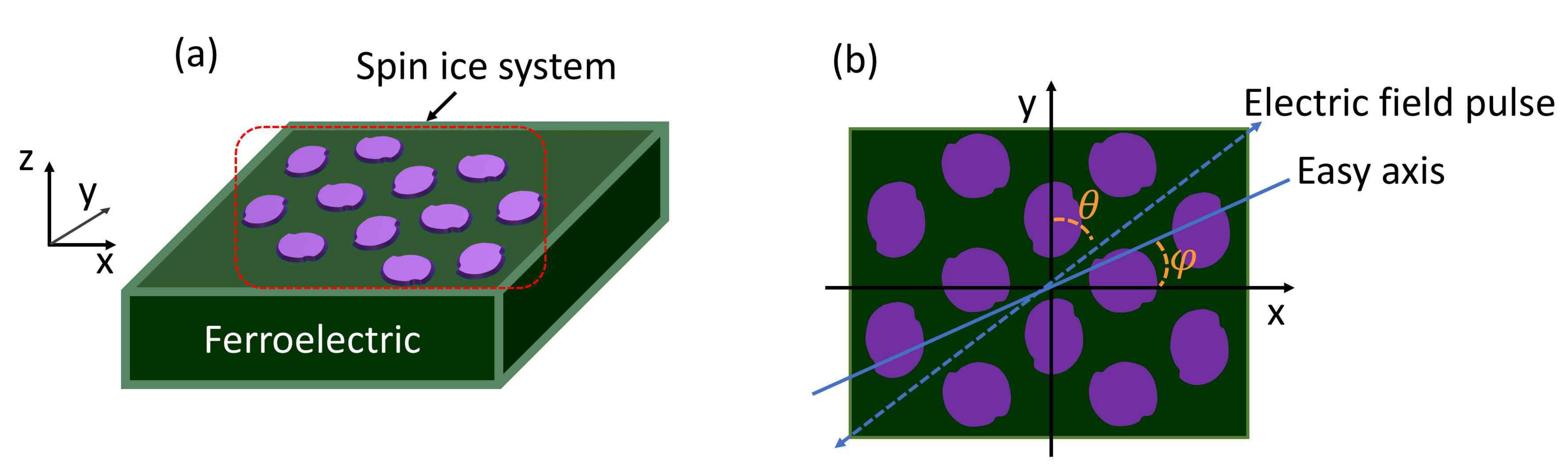}
\caption{(a) Schematic of the ASI/ferroelectric-based artificial multiferroic heterostructure device. The top view of the system is depicted in (b). Here $\theta$ is the angle between the direction of the applied electric field pulse (dashed line) and $y$-axis. The angle between the easy axis (solid line) and the $x$-axis is $\phi$ with the $x$-axis. }
\label{figure1}
\end{figure}

We design a multiferroic device utilizing piezoelectric as a  substrate and patterned peanut-shaped dipolar coupled system as a ferromagnet to demonstrate the electric field controlled magnetization switching in an artificial spin ice system as depicted in Fig.~(\ref{figure1}). We have shown the proposed peanut-shaped square ASI system coupled with a ferroelectric substrate in Fig.~\ref{figure1}(a). The top view of the designed system is shown in Fig.~\ref{figure1}(b). Here $\theta$ is  the angle  between the applied electric field pulse (dashed line) and $y$-axis; the easy axis (solid line) is making an angle $\phi$ with the $x$-axis [please see Fig.~\ref{figure1}(b)]. 
In such a system, the magnetization of the patterned nanomagnet array stabilizes along with one of the easy axes determined by the demagnetization or the dipolar field/shape anisotropy (termed as configuration anisotropy).

The effective magnetization state of such a dipolar coupled system can be controlled through electric field mediated piezoelectric strain. In such a case, a considerable anisotropic in-plane strain is produced in the piezoelectric  upon applying an electric field. The strain is transfered from piezoelectric to the ferromagnetic spin ice system, which helps in manipulating the effective magnetization of the dipolar coupled system.
Depending upon the polarity and the strength of the applied electric field, the device undergoes compressive to tensile strain resulting in different magnetization states of a dipolar coupled system through magnetoelectric effect. It is essential to consider the required energy terms such as exchange energy, energy due to the stress anisotropy, magnetostatic energy and Zeeman energy (due to an external magnetic field) to study the magnetization behaviour of such a system. Therefore, the total energy of the nanomagnet with volume V can be written as~\cite{salehi2017}
\begin{equation}
\begin{aligned}
	E^{}_i={} &\oint \bigg[A\{(\nabla m_x)^2+(\nabla m_y)^2+(\nabla m_z)^2\}+\big\{B^{}_1(\alpha^{2}_1\epsilon_{xx}+\alpha^{2}_2\epsilon_{yy}+\alpha^{2}_3\epsilon_{zz})+\\
	&B^{}_2(\alpha^{}_1\alpha^{}_2\epsilon_{xy}+\alpha^{}_2\alpha^{}_3\epsilon_{yz}+\alpha^{}_1\alpha^{}_3\epsilon_{xz})\big \}-
	\frac{1}{2}\mu^{}_o\vec{H}^{}_d\cdot\vec{M}-\mu^{}_o\vec{M}^{}\cdot\vec{H}
	\bigg]dV
	\label{eq1}
\end{aligned}
\end{equation}
In Eq.~(\ref{eq1}), the first term on the right-hand side represents the exchange energy having exchange constant A. The second term is the magnetoelastic energy, having magnetoelastic coupling constants $B^{}_1$ and $B^{}_2$. Here $\alpha^{}_i$ is the direction cosines and $\epsilon$ is the strain. The third term is the magnetostatic energy of the nanomagnet, while the final term represents the energy of interaction with an external magnetic field, $H$. In the present work, magnetocrystalline anisotropy is neglected as the nanomagnet is assumed to have random polycrystalline orientation~\cite{wang2014}. 

The dipolar field $\mu_o \vec{H}^{}_d$ shown in Eq.~(\ref{eq1}) is given by the following expression~\cite{anand2016,miltat2007}
\begin{equation}
\mu^{}_o\vec{H}^{}_d=-\frac{1}{4\pi}\int_{V}\vec{\nabla}\cdot{ \vec{M}(r')\frac{\vec{r}-\vec{r'}}{|\vec{r}-\vec{r'}|^{3}}}d^3r+\frac{1}{4\pi}\int_{S}\hat{n}\cdot{ \vec{M}(r')\frac{\vec{r}-\vec{r'}}{|\vec{r}-\vec{r'}|^{3}}}d^2r    
\label{dipolar}  
\end{equation}
Here $\hat{n}$ is the surface normal. $\vec{r}$ and $\vec{r'}$ are  the position vectors of two different nanomagnets. The dipolar field is long-range and anisotropic in nature~\cite{anand2019}. As a consequence, it drastically affects the systematic properties in such a system.

The magnetization dynamics of a nanomagnet under the influence of an effective field is described by the Landau-Lifshitz-Gilbert (LLG) equation~\cite{anand2016,arora2021}

\begin{equation}
\frac{d\vec{M}(t)}{dt}=-\gamma\vec{M}(t)\times\vec{H}^{i}_{\mathrm {eff}}(t)-\frac{\alpha\gamma}{M^{}_s}\big[\vec{M}(t)\times\vec{M}(t)\times\vec{H}^{i}_{\mathrm {eff}}(t)\big].
\label{llg}
\end{equation}
Here $\vec{H}^{i}_{\mathrm {eff}}$ is the effective magnetic field acting on the nanomagnet defined as~\cite{anand2019,arora2021}
\begin{equation}
\vec{H}^{i}_{\mathrm{eff}}=-\frac{1}{\mu^{}_oV}\frac{\partial E_i(t)}{\partial \vec{M}^{}_i(t)}
\end{equation}
$\gamma$ is the gyromagnetic ratio, $M^{}_s$ is the saturation magnetization and $\alpha$ is the damping factor associated with internal dissipation in the magnet owing to the magnetization dynamics. $E^{}_i$ the total energy of system given by Eq.~(\ref{eq1}).

The energy associated with stress anisotropy is given by~\cite{salehi2017}
\begin{equation}
E_{{\mathrm {anisotropy-stress}}}(t)=-\frac{3}{2}\lambda^{}_s\sigma(t) V\cos^2 \theta
\label{stress}
\end{equation}
where $\lambda^{}_s$ is the magnetostrictive coefficient of the magnetic material. $\sigma$ is the stress and $\theta$ is the angle between the external electric field pulse
and the $y$-axis. In this case, the effective field due to the stress anisotropy [Eq.~(\ref{stress})] takes the following form
\begin{equation}
H_{{\mathrm {eff}}-\sigma}(t)=-\frac{3}{\mu^{}_oM^{}_s}\lambda^{}_s\sigma(t) \cos^2 \theta
\label{field_stress}
\end{equation}
The effective magnetic field $H_{{\mathrm {eff}}-\sigma}(t)$ is related to an electric field of strength ${\mathcal{E}}$ as
\begin{equation}
H_{{\mathrm {eff}}-\sigma}(t)=\frac{3\lambda^{}_sY}{\mu_{o}M^{}_s(1+\nu)}(d^{}_{31}-d^{}_{32}){\mathcal{E}}
\label{electric}
\end{equation}
Here $Y$ is the Youngs' modulus; $\nu$ is Poisson ratio;  $d^{}_{31}$ and $d^{}_{32}$ are the piezoelectric charge coefficients.

We have performed Object Oriented MicroMagnetic Framework (OOMMF) based micromagnetic simulation to simulate the  magnetic moment interactions and magnetization switching dynamics in the ASI system consisting of the peanut-shaped nanomagnet.
OOMMF is based on the continuum theory of micromagnetics, which describes the magnetization process within ferromagnetic material~\cite{donahue1999}. 
These OOMMF simulations perform time integration of the LLG equation, where the adequate energy includes the exchange, anisotropy, self-magnetostatic and external magnetic fields.
In micromagnetic simulations for a single electric field pulse, the magnetization switching is observed by applying the stress of $\sigma=0.4$ GPa. The applied stress of 0.4 GPa is equivalent to a strain of $\epsilon=4\times10^{-3}$. In literature, it has been demonstrated that the strain of $1\times10^{-3}$ is equivalent to an applied electric field of 2 MV/m (20 kV/cm) across the pair of
electrodes on top of a piezoelectric (PZT) layer~\cite{cui2013}. 
Therefore, the electric field required to induce the strain of $4\times10^{-3}$ is calculated using linear interpolation for our simulations. Thus, the estimated electric field value to generate the given strain is 8 MV/m. Similarly, the strain of $2.4\times10^{-4}$ is required to switch the magnetization in a two-pulse approach, which is equivalent to 4.8 MV/m. The discretized cell size used in the simulations is 1000 nm $\times$1000 nm$\times$5 nm, implemented in the cartesian coordinate system. We have used permalloy for our study. The corresponding parameters used in the present work are: exchange constant $A =9\times10^{-12}$ Jm$^{-1}$, saturation magnetization $M^{}_s=8.60\times10^5$ $A\mathrm{m^{-1}}$, magnetocrystalline anisotropy constant $K = 0$, damping coefficient $\alpha= 0.02$, magnetostriction coefficient $\lambda^{}_s= 6\times10^{-6}$, and Young’s modulus = 100 GPa~\cite{li2007}.

\section{Simulation Results}
\begin{figure}[!htb]
\centering\includegraphics[scale=0.120]{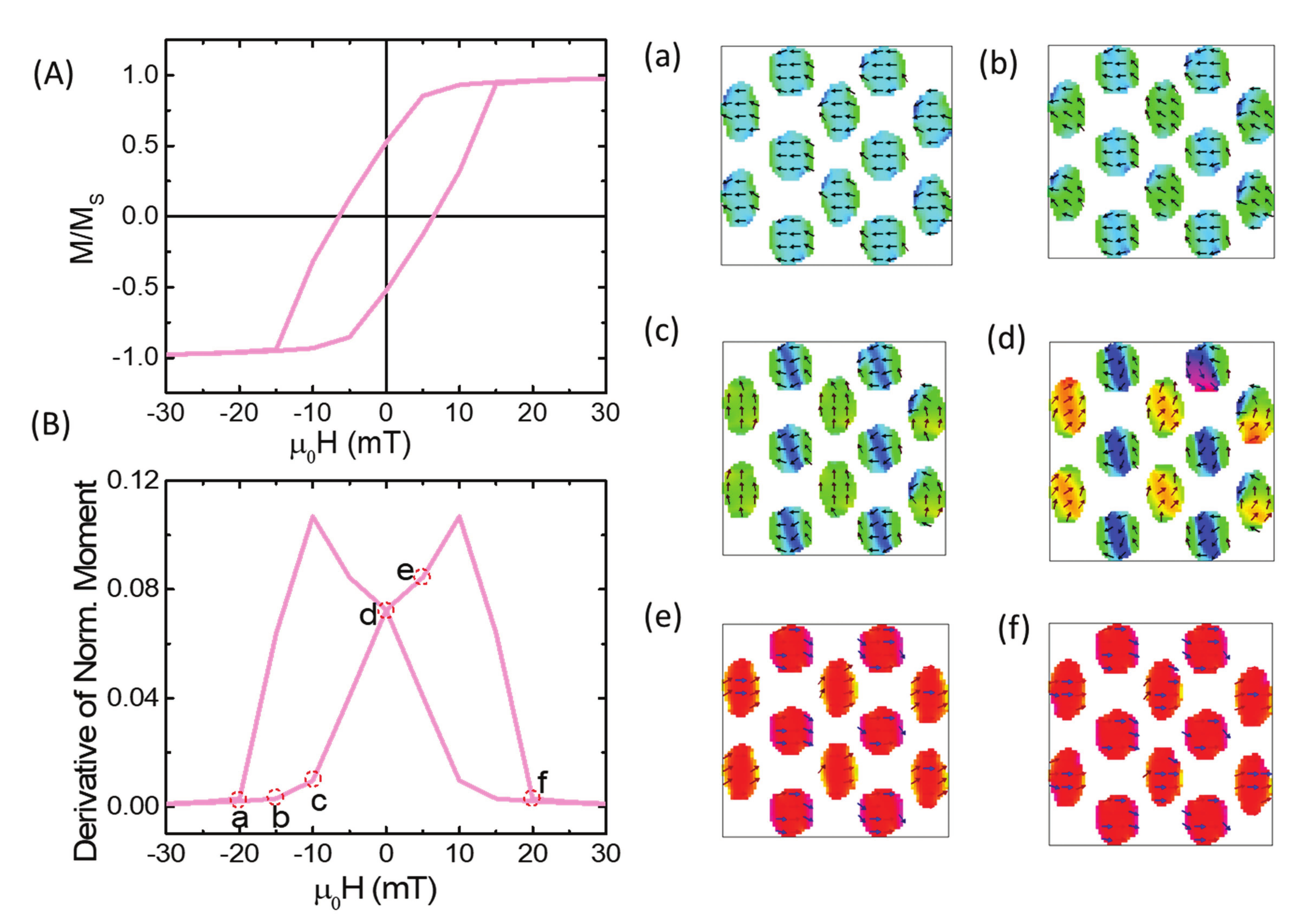}
\caption{(A) Magnetization as a function of external magnetic field and (B) Variation of the rate of magnetization change with the magnetic field. The magnetization configurations at various representative fields [marked as a, b, c, d, e, and f in (B)] are shown in (a)-(f). The 2-in/2-out magnetic state (Type II) is clearly evident at remanence [see (d)].  The other possible microstates corresponding to Type II state is also accessed by suitable value of external magnetic field}
\label{figure2}
\end{figure}

\begin{figure}[!htb]
\centering\includegraphics[scale=0.10]{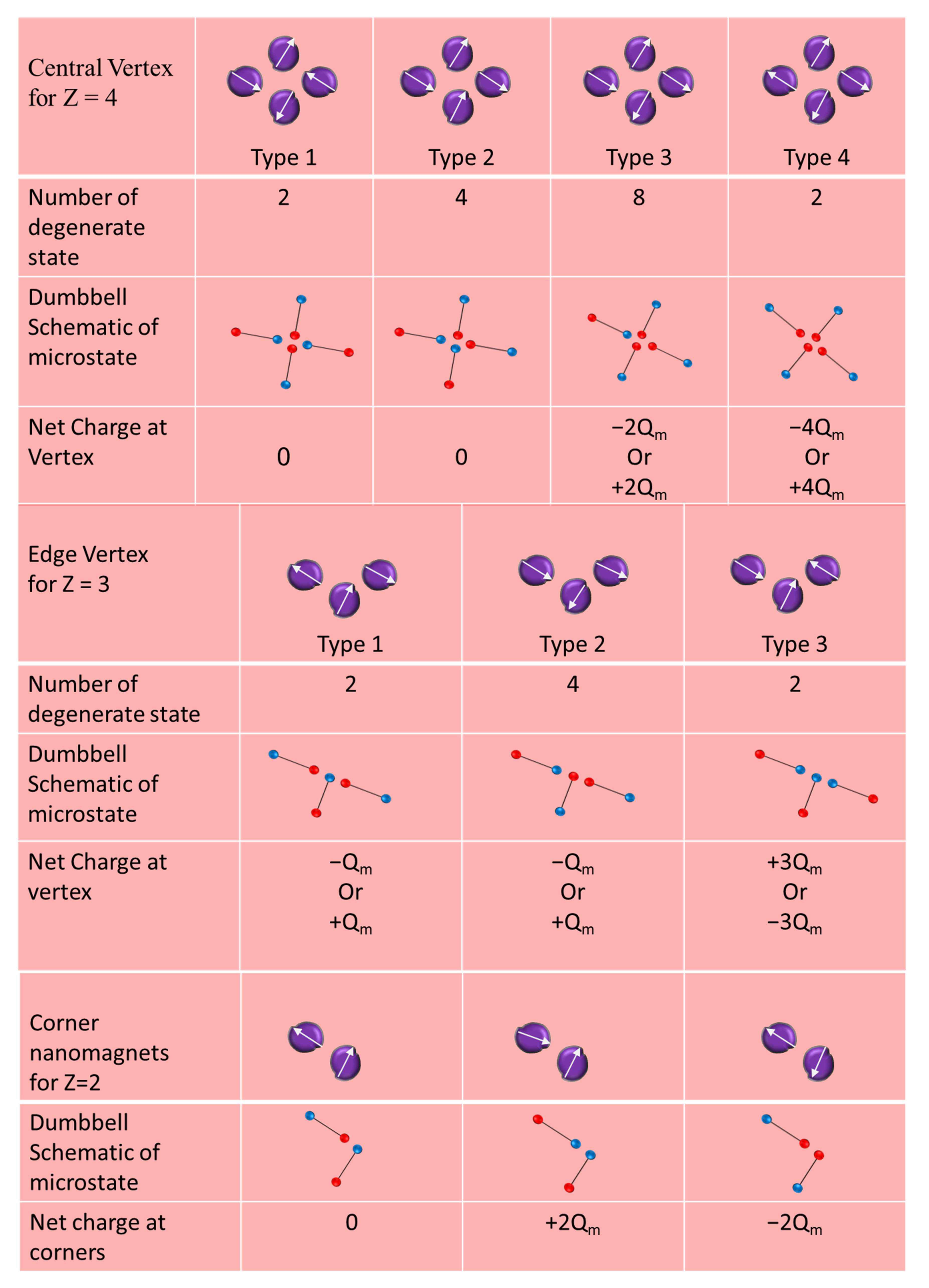}
\caption{Schematics of the spin states at the edges with $Z= 3$ and $Z= 2$.  Here, $Z$ is defined as the total number of the nanomagnets involved at the vertex edge and corner of ASI system. The corresponding degeneracy and the charge at the vertex edges for all possible state are shown as defined by dumbbell model.}
\label{figure3}
\end{figure}

 As the switching field is one of the essential quantifiers in such a system, we first investigate the magnetization switching characteristics as a function of the external magnetic field. 
The magnetic hysteresis curve of the  underlying ASI system is shown in Fig.~\ref{figure2}(A). It is evident that the coercive field  $\mu^{}_oH^{}_c$ and  the remanent magnetization $M^{}_r$ are about 6 mT and 0.5, respectively. It is also clearly seen that the anisotropy field $\mu^{}_oH^{}_K$ is about 30 mT which is very small compared to the one for highly anisotropic system such as  ASI consisting of elliptical-shaped nanomagnets ($\mu^{}_oH^{}_K\approx200$ mT)~\cite{keswani2018,keswani2019}.  We also plot the rate of change of magnetization as a function of the magnetic field to see the distinct intermediate state at various representative magnetic fields, marked as a, b, c, d, e and f in Fig.~\ref{figure2}(B).  
The analysis of these magnetic states could provide a firm basis to assess various properties of the ASI system. Fig.~\ref{figure2}(a)-(f) show them at various representative magnetic fields. The number of significant jumps in the field-dependent magnetization curve is six. 
The first magnetization switching occurs at -10 mT. There are several exciting things to note at this point [Fig.~\ref{figure2}(c)]: (1) Interestingly, there is an emergence of 2-in/2-out (Type II) magnetic configuration. (2) Four onion-type structure also emerges at the edges as depicted in Fig.~\ref{figure2}(c). (3) Remarkably, the 2-in/1-out or 2-out/1-in magnetic state emerges at the four vertices with $Z = 3$ at the edges are naturally magnetically charged with $+Q^{}_m$ or $-Q^{}_m$ [please see the schematic Fig.~(\ref{figure3}) for reference]. (4) Likewise, the four corners with $Z=2$ have an absolute charge of $2Q^{}_m$ or zero [see Fig.~(\ref{figure3}) for clarity]. Therefore, the total magnetic charge in the system is zero. It implies that the system maintains magnetic charge neutrality even after the magnetization of individual nanomagnets gets flipped because of an external magnetic field. This magnetization switching is dominated by dipolar interaction, corresponds to  demagnetization energy $\approx5.07\times10^{-17}$ J, which is one order larger than the exchange energy ($=3.07\times10^{-18}$ J) counterpart. The next jump occurs at the coercive field $\approx6$ mT, 2-in/2-out (Type II) magnetic state persists at the central vertex as shown in Fig.~\ref{figure2}(d). The emergence of such a complex magnetic state can also be attributed to the demagnetization field as the corresponding energy $\approx 4.69\times10^{-17}$ J is dominant as compared to the exchange energy $\approx2.86\times10^{-18}$ J. One can draw similar observations at other magnetic fields. 
Notably, the coherent switching of the vertically and horizontally aligned nanomagnets indicate as if they are locked. Such magnetization switching in unison like coupled systems is primarily because of demagnetization interaction. Consequently, the horizontally placed nanomagnets change their magnetization in unison. Likewise, vertically aligned nanomagnets change their magnetization together.

\begin{figure}[!htb]
\centering\includegraphics[scale=0.10]{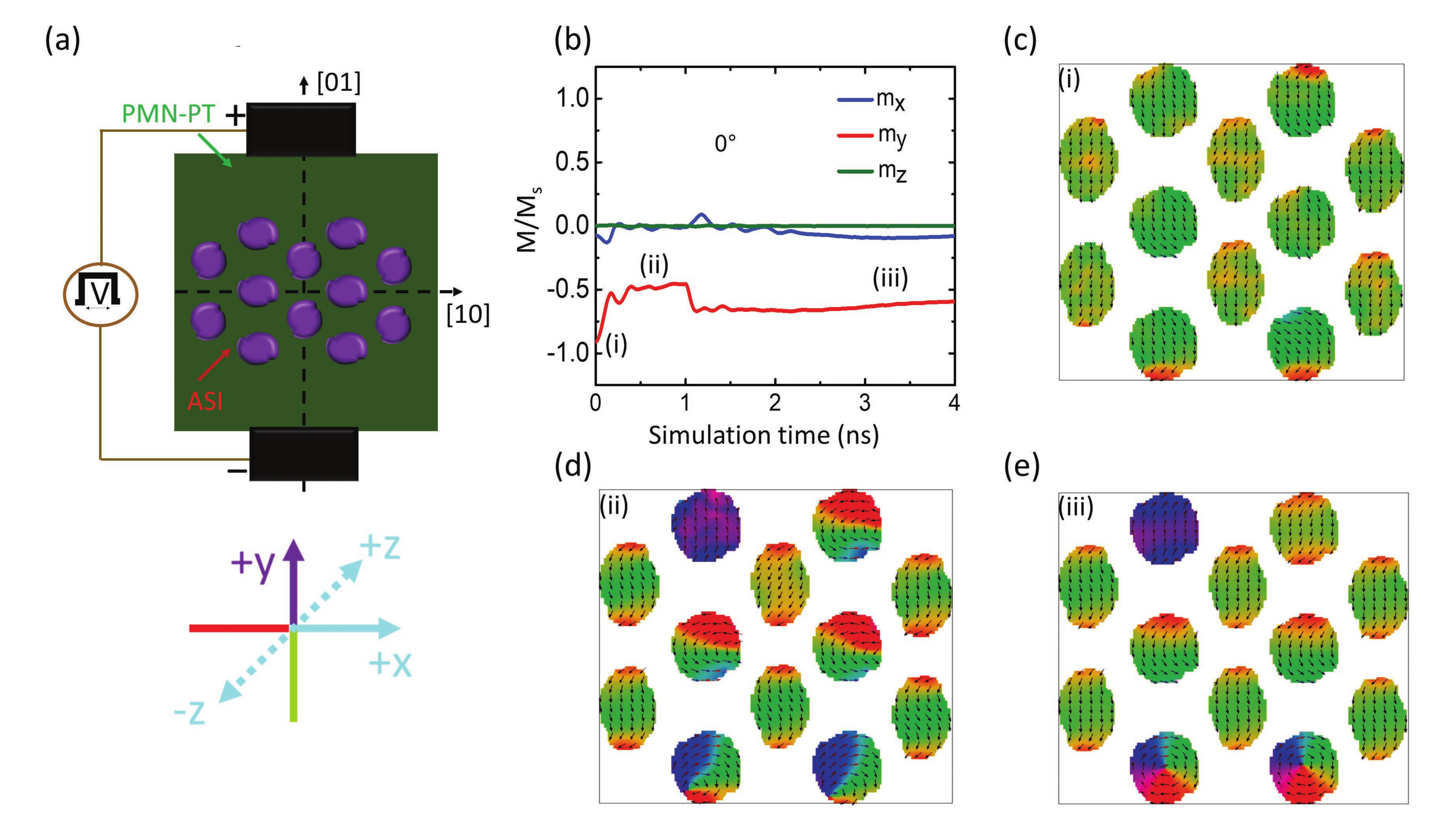}
\caption{(a) Schematic of the device by applying the electric field pulse at $\theta= 0^\circ$. (b) The time evolution of magnetization upon application of an electric field pulse (c) Initial magnetization state (d) intermediate magnetization state (e) final magnetization state of the ASI system.} 
\label{figure4}
\end{figure}

 The electric field pulse of 8 MV/m strength with the pulse width of 1 ns (equivalent to stress of 0.4 GPa) is applied along the $y$-axis, i.e. along [0 1] direction with $\theta = 0^\circ$ to probe the magnetization switching dynamics due to the piezoelectric strain, as shown in schematic  Fig.~\ref{figure4}(a). Fig.~\ref{figure4} (b) shows the time evolution of the magnetization component along $x$, $y$ and $z$ directions for the electric field pulse applied along the $y$-axis. Intially, the magnetization is along $-y$ direction as shown in Fig.~\ref{figure4}(c). (a) Upon application of the electric field pulse, the magnetization precesses and reaches from an initial state (i) at $t= 0$ ns to an intermediate state (ii) in 0.5 ns [see Fig.~\ref{figure4}(c) and (d)]. After switching off the external electric field, the intermediate magnetic  state relaxes to the final state (iii), similar to initial state (see Fig.~\ref{figure4} (e)) as most vertically oriented nanomagnets seem to revert to similar colour distribution. However, some of the horizontally oriented nanomagnets seem to have significantly different colour distributions, i.e. different magnetization states and hence we can conclude that only partial magnetization switching is observed by electric field pulse applied along the [0 1] direction ($y$-direction).

\begin{figure}[!htb]
\centering\includegraphics[scale=0.12]{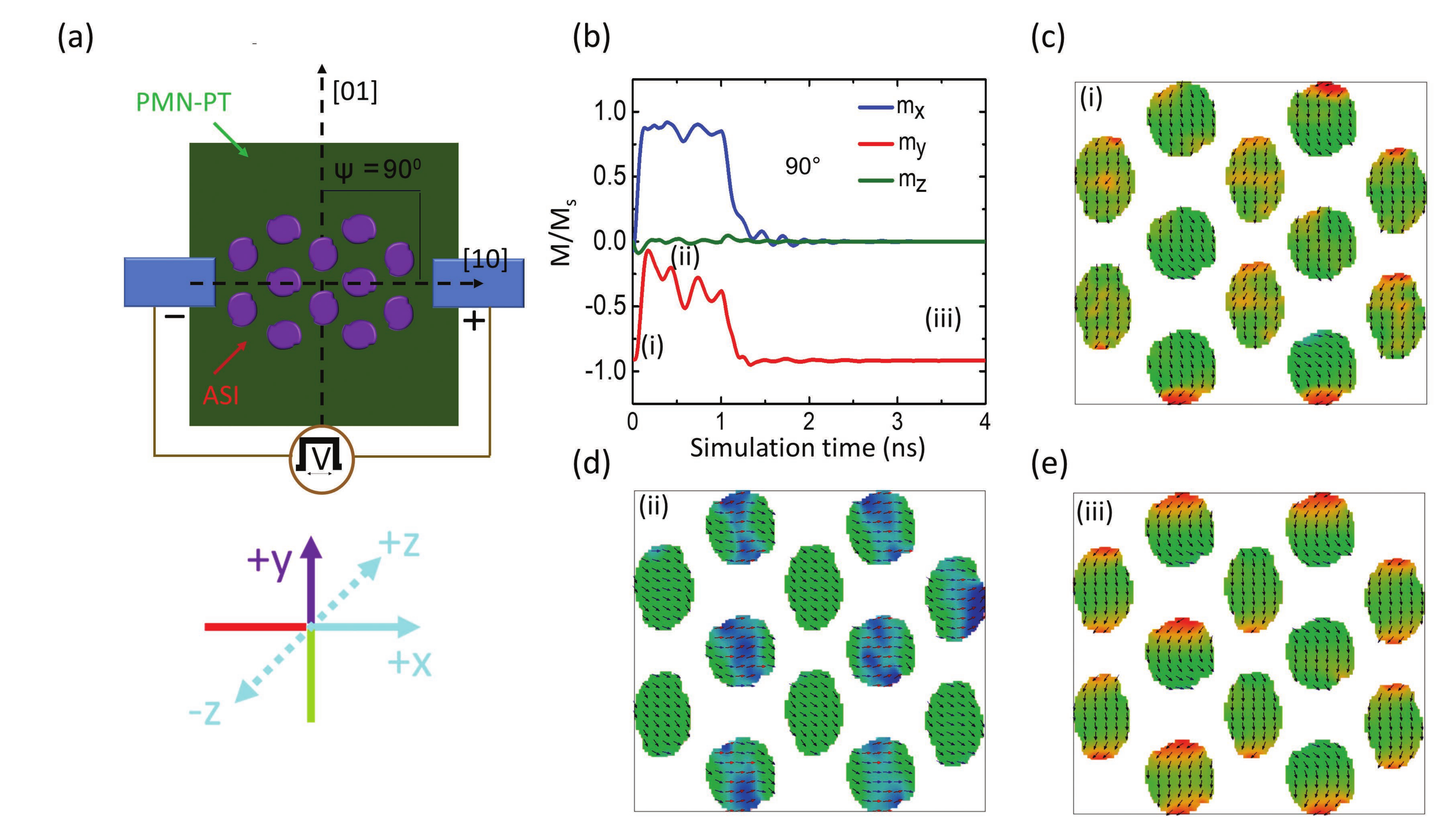}
\caption{(a) Schematic of the device with the electric field pulse applied at $\theta= 90^\circ$. (b) Time evolution of magnetization along by applying electric field pulse along $x$-axis. (c) Initial magnetization state, (d) intermediate magnetization state, and (e) final magnetization state of the ASI system.}
\label{figure5}
\end{figure}

Further, the time evolution of the magnetization of an ASI system is investigated by applying the electric field pulse of 8 MV/m at $\theta=90^\circ$ to the $y$-axis of the ASI
system, i.e. along [1 0] direction, as shown in Fig.~\ref{figure5}(a). Fig.~\ref{figure5}(b) shows the time evolution of the magnetization along the $x$, $y$ and $z$-axis upon application of electric field pulse. The magnetization switching and corresponding magnetic states at initial ($t = 0$ ns), intermediate ($t = 0.5$ ns), and final stage (after $t = 3$ ns) marked as (i), (ii) and (iii) in Fig.~\ref{figure5}(b), are shown in Fig.~\ref{figure5}(c)-(e), respectively. It is clearly seen that the magnetization is aligned along the negative $y$-direction initially. The magnetization starts to get aligned along the $x$-direction upon application of electric field pulse. As the pulse is switched off, the magnetization gets back to its
initial orientations (along the $-y$-axis), as indicated in Fig.~\ref{figure5}(e). These observations can be explained from the energy barrier approach as follows. The natural tendency of the magnetization is to get aligned along in the anisotropy direction, which is along the diagonal axis of the system ($\sim45^\circ$). 
While the magnetization is forced to get aligned along the $x$-axis ($\sim45^\circ$ away from the easy axis). The energy cost to remain in this state is very large; therefore, effective
magnetization tends to find an escape route to retain its initial state by finding the nearest energy minimum once the electric-field pulse or the strain is removed. So, it retains the initial state, i.e. no partial switching, which is a stable state of the magnetization.

\begin{figure}[!htb]
\centering\includegraphics[scale=0.12]{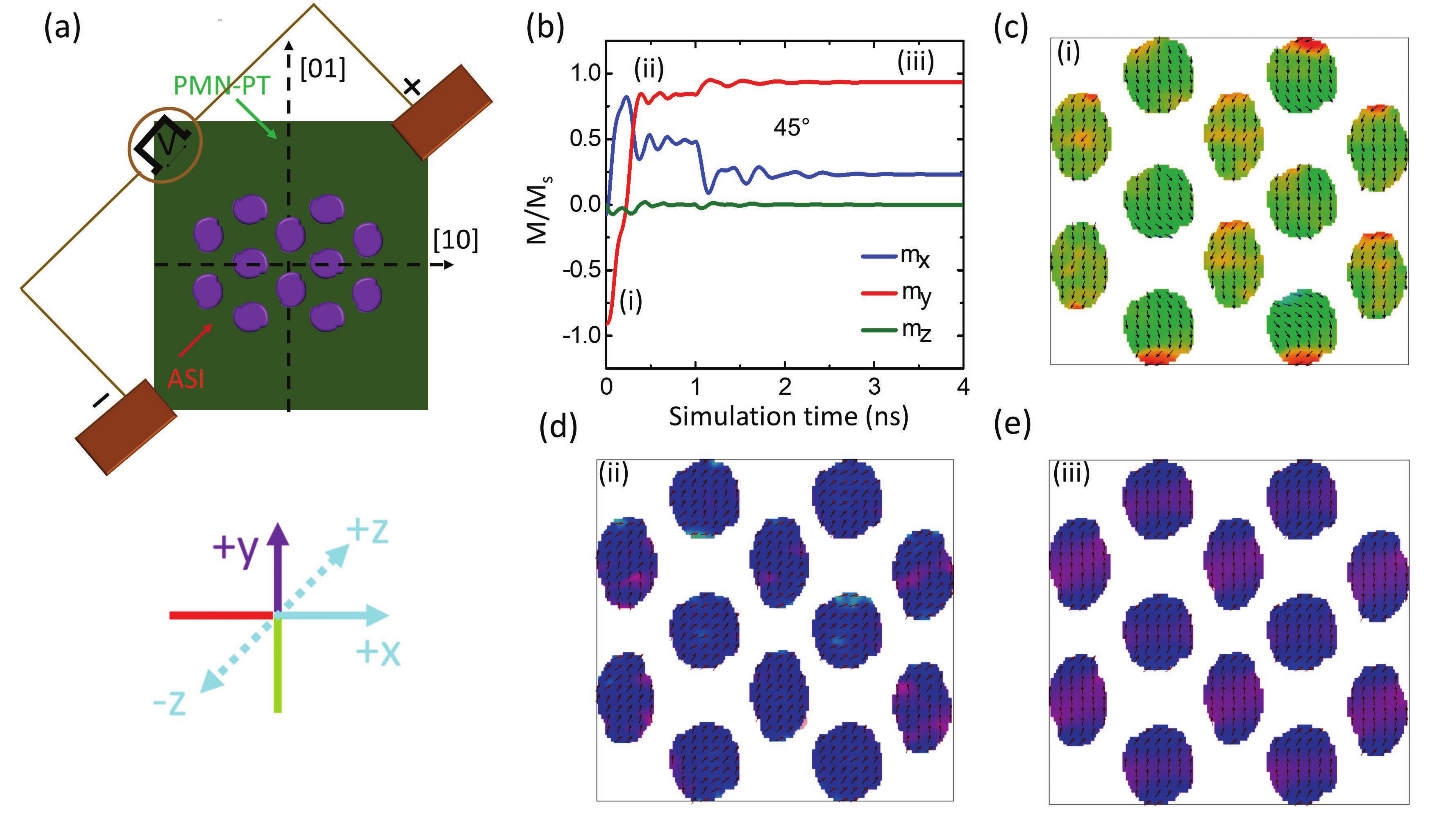}
\caption{(a) Schematic of the device with the electric field pulse at $\theta = 45^\circ$. (b) Time evolution of magnetization along $x$, $y$ and $z$ direction by applying electric field pulse (c) Initial magnetization state (d) intermediate magnetization state (e) final magnetization state of the ASI system.}
\label{figure6}
\end{figure}

\begin{figure}[!htb]
\centering\includegraphics[scale=0.12]{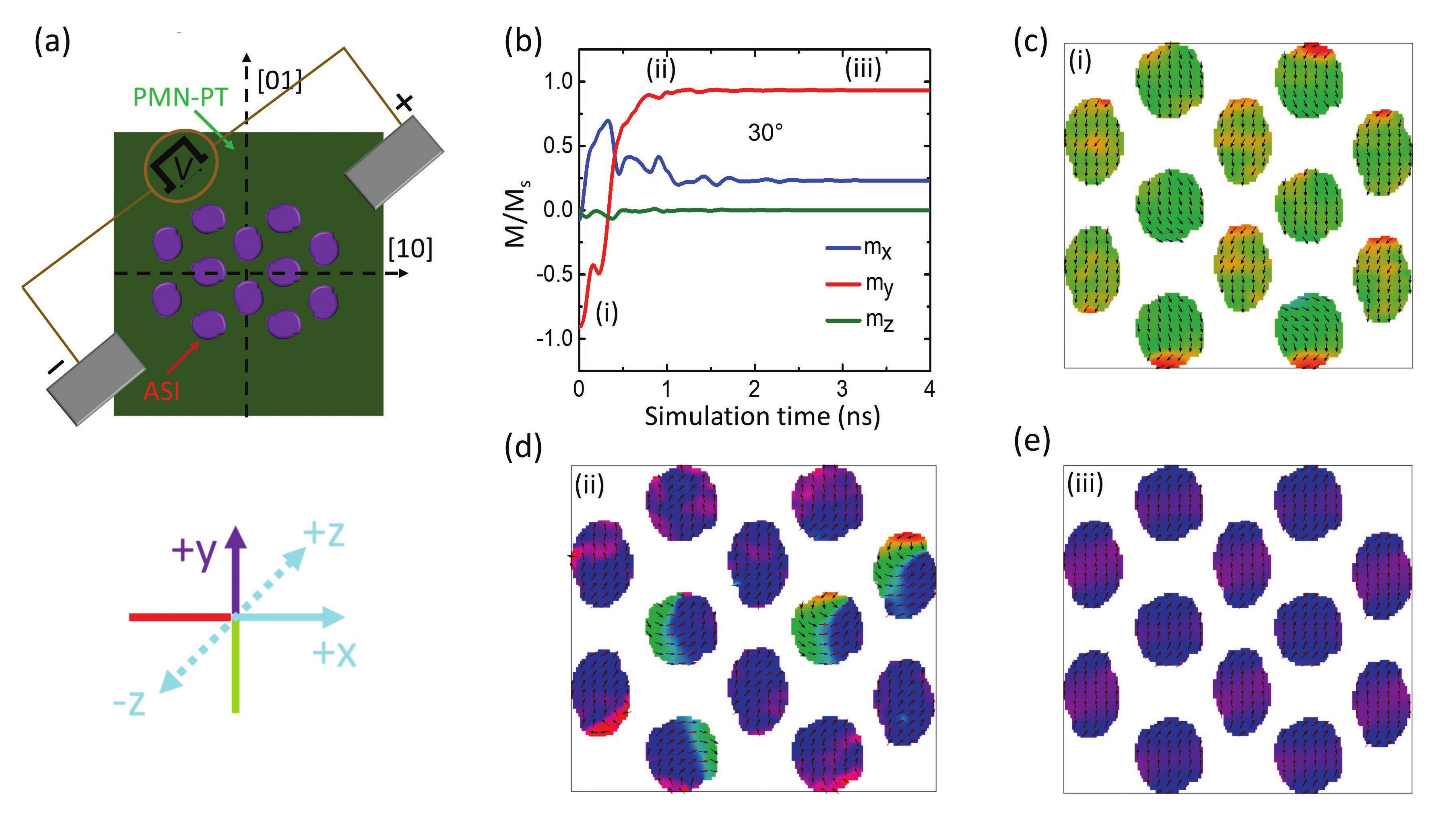}
\caption{(a) Schematic of the device with the electric field pulse applied at $\theta = 30^\circ$. (b) Time evolution of magnetization along $x$, $y$ and $z$ direction by applying electric field pulse (c) Initial magnetization state (d) intermediate magnetization state (e) final magnetization state of the ASI system.}
\label{figure7}
\end{figure}

The observation of either no or partial magnetization switching in the ASI system with the electric field pulse applied along the $x$ and $y$-axis, respectively suggests that we can achieve the complete magnetization reversal by changing the direction of the applied electric field pulse. Therefore, we now study the magnetization switching dynamics by using the electric field pulse of 8 MV/m at different angles, $\theta$, in the range of $0^\circ$ to $180^\circ$. The study of time-dependent magnetization of the underlying system by applying the electric field pulse of strength 8 MV/m along the $45^\circ$ from the positive $y$-axis is studied and shown in Fig.~\ref{figure6}(a). Once again, the initial magnetization is set to negative y-direction, as shown in Fig.~\ref{figure6}(c) at
time $t = 0$ ns. As the electric field pulse is applied, the magnetization is forced to align along the positive $x$-direction, as indicated by the rise of $m^{}_x$ in Fig.~\ref{figure6}(b).
As the electric field pulse is switched on completely at  $t\approx0.25$ ns, there is a complete and coherent magnetization reversal in all the constituent nanomagnets. Remarkably, the magnetization of the system switches mostly along positive $y$-direction at intermediate state position ($t = 0.5$ ns), as indicated by predominantly $m^{}_y$ signal. The magnetization remains stabilized along the $+y$ even after removing the external field pulse [see Fig.~\ref{figure6}(e)]. Hence, we can switch the magnetization by $180^\circ$ (from $-y$ to $+y$ direction) by applying the electric field pulse at angle $\theta=45^\circ$. Similarly, the complete magnetization reversal can also be realized by applying the electric field pulse of similar strength at $\theta=30^\circ$, as shown in Fig.~(\ref{figure7}). Interestingly, there is partial magnetization switching at the intermediate state, as evident in Fig.~\ref{figure7}(d). Notably, the magnetization finds the nearest minimum along the $+y$ direction upon removal of the electric field pulse, evident from very large $m^{}_y$ signal [see Fig.~\ref{figure7}(e)], indicating the complete magnetization reversal.

\begin{figure}[!htb]
\centering\includegraphics[scale=0.12]{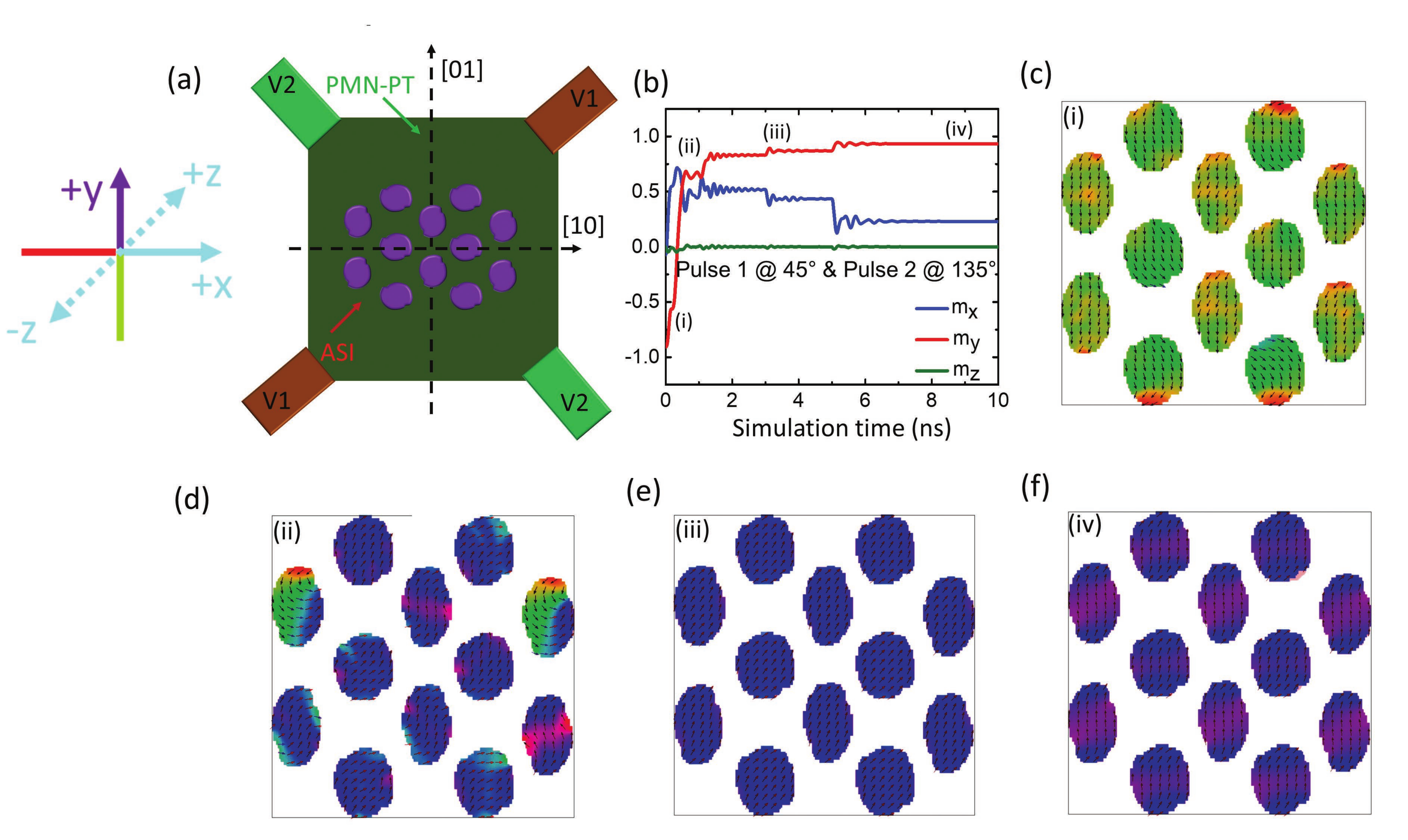}
\caption{(a) The schematic of the device with the applied electric field pulses at $\theta = \pm45^\circ$ with respect to $+y$ direction, (b) The variation of the magnetization with simulation time of the system along $x$, $y$ and $z$ direction with the sequential applied electric field pulse along $\theta= \pm45^\circ$. (c) Initial state of the magnetization (d) first intermediate magnetization after the pulse applied at $-45^\circ$ (e) Second intermediate state after the pulse applied at $+45^\circ$. The stable state of magnetization state after the removal of electric field pulse.} 
\label{figure8}
\end{figure}

\begin{figure}[!htb]
\centering\includegraphics[scale=0.12]{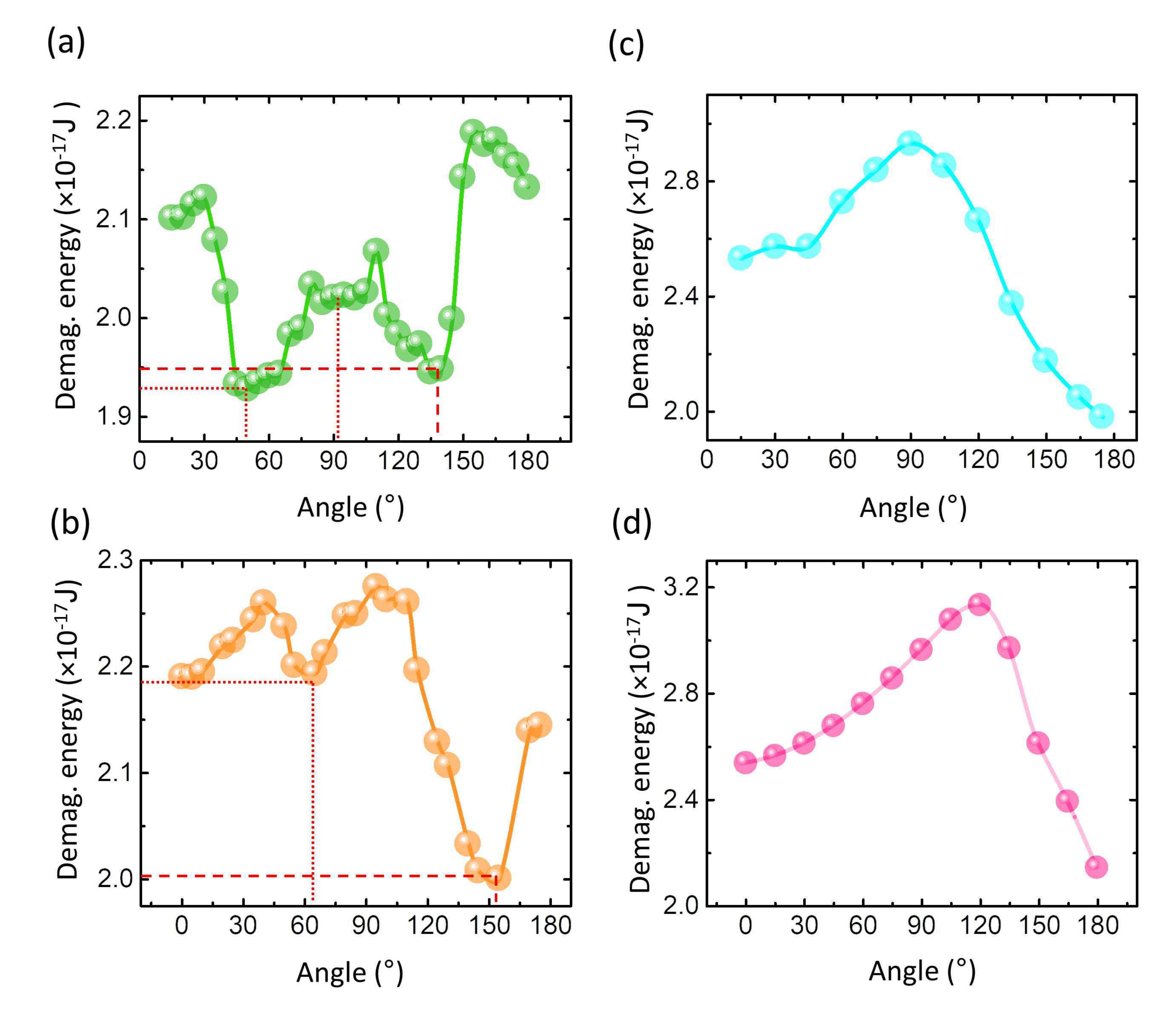}
\caption{The variation of demagnetization energy with the rotation of applied magnetic field at different strain states (a) 0 MV/m (b) 1.6 MV/m (c) 4.8 MV/m (d) 8 MV/m.} 
\label{figure9}
\end{figure}

Further, we explored another scheme for magnetization switching utilizing two sequential electric field pulses of smaller magnitude than a single field pulse. It could be beneficial for materials (used for developing ASI systems) with a smaller magnetostriction coefficient. In this scheme, the sequential electric field of $4.8$ MV/m (relatively smaller strength compared to single field pulse of strength 8 MV/m) of 1 ns each separated by 1 ns is applied at two pairs of electrodes at $\pm45^\circ$ with respect to the $+y$-axis as shown in Fig.~\ref{figure8}(a). The evolution of magnetization
with time is shown in Fig.~\ref{figure8}(b). As before, the initial magnetization is set to negative $y$ direction [see Fig.~\ref{figure8}(c)]. As the first pulse is applied at $-45^\circ$, the magnetization is forced to align initially in the $+x$ direction, as indicated by the sharp rise in the $m^{}_x$ component. It tends to get aligned  along the system's diagonal direction, which is one of the stable states of the dipolar coupled system at intermediate state (ii) (see Fig.~\ref{figure8}(d)). 
Consequently, the magnetization predominantly switches in all peanut nanomagnets coherently (see Fig.~\ref{figure8}(e)) along the $+y$ axis as inferred from the very large $m^{}_y$ and decreased $m^{}_x$ at state (iii). Remarkably, the magnetization stays and gets stabilized along $+y$-direction at last state (iv). The above results clearly indicate that  we have achieved complete magnetization reversal through the sequential electric field pulses induced strain of lower magnitude. 
In order to probe the angle-dependent magnetization switching due to electric field pulse, 
we extensively investigate the variation of the energy barrier as a function of an external magnetic field to determine the natural  orientation of the anisotropy (shape). The anisotropy energy $ E$ can be expressed as~\cite{carrey2011}
\begin{equation}
E=E^{}_b\sin^2\delta.
\label{barrier}
\end{equation}
Here $E^{}_b$ is the energy barrier seen by the magnetic moment, and $\delta$ is the angle between the anisotropy axis and the magnetization of a dipolar coupled system. As we have used polycrystalline material in the present work, magnetocrystalline anisotropy is negligibly small, i.e. $K = 0$. Therefore, the main contribution to the energy barrier should come from the demagnetization energy. As it is also a well-known fact that the main reason for shape anisotropy is the
demagnetization field, Eq.~(\ref{barrier}) can therefore be used to determine the anisotropy direction correctly. We calculated the demagnetization energy at different angles of rotations of the constant applied magnetic field for our ASI system comprising peanut-shaped NiFe nanomagnets coupled to ferroelectric piezoelectric substrate in the absence and the presence of an external electric field, as shown in Fig.~(\ref{figure9}). In the absence of applied electric field pulse [see Fig.~\ref{figure9}(a)], two distinct energy minima separated by an energy barrier are observed. Such an energy barrier is primarily due to the demagnetization interaction of the underlying system. These two minima are found roughly at about $45^\circ$ and $135^\circ$, separated by $90^\circ$ [see Fig.~\ref{figure9}(a)], with the energy barrier hump centred at about $90^\circ$. Note that the demagnetization energy value at the two minima is almost similar. Therefore, It clearly implies that the angle between our ASI system's easy axis and hard axis is about $45^\circ$, marked with an arrow, depicted in Fig.~\ref{figure9}(a). After that, we analyze the demagnetization energy variation in the presence of the electric field. In Fig.~\ref{figure9}(b)-(d), the variation of the demagnetization energy is investigated by applying the electric field of strength varied between 1.6 MV/m to 8
MV/m. Remarkably, the two energy minima shift to $60^\circ$ and $150^\circ$ (still separated by $90^\circ$), with the barrier peak approximately in-between at about $105^\circ$ for the electric field strength $=1.6$ MV/m. However, there is a significant difference: the energy minima are now asymmetric about the barrier peak with the energy minima on one side (the one at about $60^\circ$) of the barrier lifts to a higher value than the other side (at about $150^\circ$) of the barrier. It will result in one direction of the magnetic moment (which are at energy minima of $60^\circ$) to get aligned in the opposite direction magnetic moments (at energy minima of
$150^\circ$) through a complete reversal. It is found that when the applied electric field-induced strength is sufficiently large enough (8 MV/m), it rotates the direction of a magnetic moment from one minimum state to another minima by crossing the energy barrier (see Fig.~\ref{figure9}(d)). As a consequence, the magnetization of the ASI system changes its direction by $180^\circ$. The above procedure is far superior in comparison to the free energy approach to find out the easy axis orientation of dipolar coupled complex nanostructures.

\section{SUMMARY AND CONCLUSION}
To summarize, we have performed micromagnetic simulations to investigate the electric field-induced strain mediated magnetization switching
in artificial spin ice (ASI) based multiferroic system. 
However, the piezoelectric strain is not able to switch the magnetization by $180^\circ$ in such a system because of its uniaxial character. Therefore, we propose an ASI system consisting of a peanut-shaped nanomagnet whose angle between the easy and hard axes is less than $90^\circ$. In such a system, the anisotropy field $\mu^{}_oH^{}_K$ comes out to be $\approx30$ mT, which is about seven times smaller as compared with highly anisotropic nanomagnets such as ellipsoid ($\mu^{}_oH^{}_K=200$ mT)~\cite{keswani2018}. Therefore, the complete magnetization reversal can be accomplished easily using field pulse of permissible strength (accordance with experiments) in our proposed system.
Remarkably, the magnetization switching depends strongly on the direction of the externally applied electric field pulse. It implies that the magnetization reverses its direction only if the electric field pulse is applied at some particular angles ($\theta\sim 30-45^\circ$), close to the anisotropy direction of the system. Our analysis of the demagnetization energy variation reveals that the energy barrier becomes antisymmetric in such cases, which facilitates the complete magnetization reversal.
Moreover, we have also proposed and successfully demonstrated an alternative way to switch the magnetization using two sequential electric field pulses of relatively smaller magnitude. The two pulse approach is highly desirable to get the complete reversal of magnetization in ferromagnetic material having a lower value of magnetostriction coefficient.
Therefore, the present work could instigate  extensive experimental, theoretical and computational research in these extraordinarily versatile and valuable systems.
We hope that our work could pave the way for future ASI-based scalable magnetic field-free low power spintronics devices.


\section*{ACKNOWLEDGMENTS}
AC would like to acknowledge the NTU research scholarship (NTU-RSS) for funding support. R.S.R. would like to acknowledge the Ministry of Education (MOE), Singapore through Grant No. MOE2019-T2-1-058 and National Research Foundation (NRF) through Grant No. NRF-CRP21-2018-0003.

\section*{DATA AVAILABILITY}
The data that support the findings of this study are available from the corresponding author upon reasonable request.
\bibliography{ref}
\newpage
\newpage

\newpage
\newpage

\newpage
\newpage

\newpage

\newpage

\newpage
\end{document}